# BARRIER LAYER FORMATION AND PTCR EFFECT IN (1-x) [Pb(Fe$_{1/2}$Nb$_{1/2}$)O$_3$]-xPbTiO$_3$ ( x = 0.13) CERAMICS

SATENDRA PAL SINGH, AKHILESH KUMAR SINGH and
DHANANJAI PANDEY
School of Materials Science and Technology, Institute of Technology,
Banaras Hindu University, Varanasi-221 005, India

(1-x)Pb(Fe$_{1/2}$Nb$_{1/2}$)O$_3$-xPbTiO$_3$ (PFN-xPT) ceramics with x = 0.13 sintered at 1200 °C show diffuse phase transition and very high dielectric constant at lower frequencies. The high value of dielectric constant at lower frequencies is shown to be due to the barrier layer formation. The resistivity of the PFN-xPT ceramics, obtained by complex impedance analysis, initially decreases with temperature and then shows an upward trend beyond the ferroelectric Curie point reminiscent of BaTiO$_3$ based thermistors with PTCR effect.

**Keywords:**   Barrier Layer; PTCR; Diffuse Ferroelectric Transition.

## INTRODUCTION:

It is well known that the donor-doped BaTiO$_3$ ceramics exhibit an anomalous increase in resistivity by 3 to 7 orders of magnitude near the ferroelectric Curie point. This behaviour is commonly known as positive temperature coefficient of resistivity (PTCR) [1]. The PTCR effect is utilized in thermistors, which have many technological applications such as temperature sensor, fluid flow and liquid level sensors and power regulators in electronic circuits. Besides doped

BaTiO$_3$, PTCR effect is also observed in several other ferroelectric materials like KNbO$_3$ [2], NaBiTi$_2$O$_6$ [3] and KBiTi$_2$O$_6$ [3]. PTCR effect has also been reported in Pb(Fe$_{1/2}$Nb$_{1/2}$)O$_3$ (PFN), which exhibits a diffuse non relaxor type transition [3,4]. Recently we [5] studied the effect of PbTiO$_3$ (PT) substitution on the structure and dielectric behaviour of PFN. It was shown that the high dielectric constant of these mixed (1-x) Pb(Fe$_{1/2}$Nb$_{1/2}$)O$_3$-xPbTiO$_3$ (PFN-xPT) ceramics at low frequencies is due to the formation of barrier layers. The barrier layer formation was shown to be independent of the pyrochlore phase content. It was also shown that with increasing PT content, or decreasing sintering temperature the barrier layer effect is diminished. In the present work, we have studied the temperature dependence of real and imaginary parts of the dielectric constant ($\varepsilon'$, $\varepsilon''$) and impedance ($Z'$, $Z''$) of PFN-0.13PT ceramics sintered at 1200 °C. It is shown that the resistivity of these ceramics exhibits negative temperature coefficient (NTC) up to the Curie point after which it shows positive temperature coefficient (PTCR effect). The redox reactions responsible for the PTCR effect are discussed.

**EXPERIMENTAL:**

Multi-step mixed oxide route was used for the synthesis of PFN-xPT powders the details of which are described elsewhere [5]. The powders were sintered at 1200 °C in closed PbO atmosphere with about 1.8% gain in weight after sintering. The pyrochlore phase content after sintering was found to be 0.6% as estimated using the relative intensities of the pyrochlore 222 and perovskite 110 peaks. For the dielectric and impedance measurements, the surface of sintered pellets was first gently polished with 0.25μm diamond paste and then silver paste was applied on both sides of the pellets. The pellets were dried at 100 °C in an oven and then cured by firing at 500 °C for about 10 minutes. High temperature dielectric measurements and impedance analysis were carried out using an LCR meter (Hioki 3532 LCR Hitester). Temperature was recorded using an Indotherm programmable temperature controller and chromel alumel thermocouple with an accuracy of ±1 °C. The impedance analysis was carried out using Z-Plot software provided by Solartron.

## RESULTS AND DISCUSSION:

Fig. 1 shows the variations of real ($\varepsilon'$) and imaginary ($\varepsilon''$) parts of the dielectric constant with temperature at various frequencies for PFN-0.13PT. It is evident from this figure that the phase transition is very diffuse and shows very strong frequency dispersion at all temperatures. However it does not correspond to a relaxor ferroelectric behaviour since the peaks in $\varepsilon'$ and $\varepsilon''$ occur at the same Curie point which is frequency independent also. For relaxors, the two peak temperaturs are non-coincident and both are frequency dependent [6,7,8]. Thus the dielectric response shown in Fig. 1 is due to a non-relaxor type diffuse ferroelectric transition.

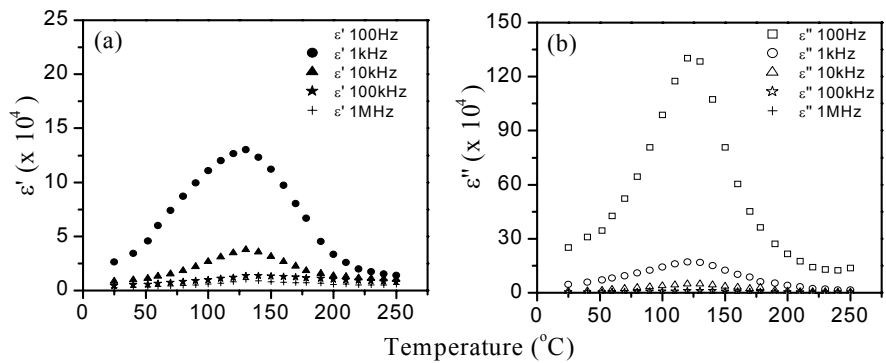

FIGURE 1, Variation of the (a) real ($\varepsilon'$) and (b) imaginary ($\varepsilon''$) parts of dielectric constant of 0.87PFN-0.13PT sintered at 1200 $^\circ$C with temperature in the frequency range 100Hz -1MHz.

The measured dielectric constant decreases to rather very low values for frequencies greater than 10kHz as shown in Fig.2 at room temperature. Such a drastic decrease in the value of the dielectric constant at higher frequencies can be explained in terms of the interfacial polarization. Interfacial polarizability results due to the difference in the conductivity of the grains and the grain boundaries. This difference in the conductivity arises due the loss of oxygen during sintering and its reintake during cooling at the grain boundaries. The semiconducting behaviour of the grains in PFN ceramics is believed to be due to the loss of oxygen during firing at higher temperatures in accordance with the reaction [9] (1) where all the species are written in accordance with Kröger Vink notation of defects. Electrons released in reaction (1), may be captured by $Fe^{3+}$ leading to the formation of $Fe^{2+}$ according to the reaction (2).

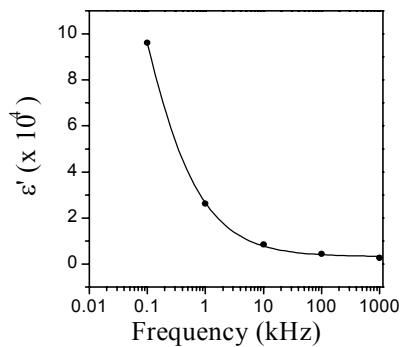

FIGURE 2, Variation of room temperature dielectric constant with frequency.

$$O_o \Leftrightarrow \frac{1}{2} O_2 + V_o^{\bullet\bullet} + 2e^{/}, \quad (1)$$

$$Fe^{3+} + e^{/} \rightarrow Fe^{2+} \quad (2)$$

This leads to hopping of electrons between the $Fe^{2+}$ and the $Fe^{3+}$ ions, and increases the conductivity [9]. During cooling after sintering, the reverse reaction of (1) occurs. But due to insufficient time available during cooling and the falling temperature, the reoxidation process is restricted to the grain boundaries only[10] which restores the insulating behaviour to the grain boundary regions only. The grain-grain boundary interface layer is usually called as 'barrier layer'. In order to confirm the formation of barrier layers in the PFN-0.13PT ceramics, we have analyzed the complex impedance using Cole-Cole impedance plots. Fig. 3 (a-f) depicts the Cole-Cole impedance plots at various temperatures below and above the Curie point. Cole-Cole plots in Fig. 3(a-e) clearly show the presence of two depressed circles. These two circles represent the contributions of grains and the grain boundaries and confirm the formation of barrier layers [10] responsible for the extremely high dielectric constant in PFN-xPT. There was no contribution from the electrode-specimen interface, as the third circle on low frequency side was not observed. The two circles in the Cole-Cole plot merge into one circle with increasing temperature as can be seen from Fig. 3(f) for 240 °C. The relaxation times for the two polarization processes representing the grain and the grain boundary contributions seem to change in different fashion because of the difference in their activation energies. However, at higher temperatures both the processes seem to have comparable relaxation time giving rise to one semicircle in the Cole-Cole plots.

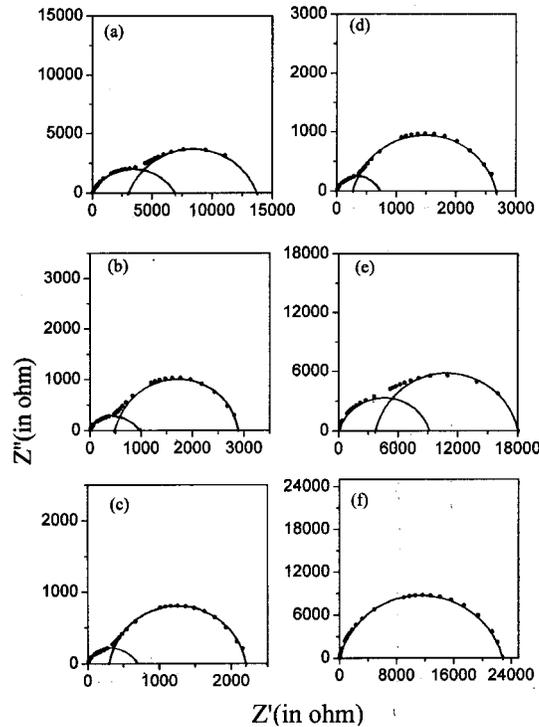

FIGURE 3, Cole-Cole impedance plots for 0.87PFN-0.13PT at various temperatures (a) 25 °C (b) 100 °C (c) 130 °C (d) 140 °C (e) 210 °C (f) 240 °C

Figure 4. shows the variation in the resistivity with temperature, as obtained from the intersection of the lower frequency semicircle with the $Z'$ axis shown in Fig. 3. The resistivity is found to decrease with increasing temperature upto 125 °C and exhibits a negative temperature coefficient of resistivity (NTC effect). Beyond the Curie point, the resistivity increases with increasing temperature and shows a positive temperature coefficient of resistivity (PTCR effect). Unlike doped $BaTiO_3$ ceramics, where the PTC resistivity jumps by 3 to 7 orders of magnitude, the effect here is rather very weak and the increase is by only one order of magnitude. This may be due to the difference in the schottky barrier height '$\phi$' [11; 12] formed at the grain boundaries for $BaTiO_3$ and PFN-0.13PT ceramics,

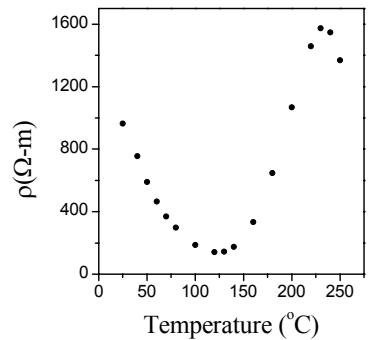

FIGURE 4, Variation of the resistivity of 0.87PFN-0.13PT with temperature.

$$\phi = \frac{N_D d^2 e^2}{2\varepsilon_r \varepsilon_o}, \qquad (3)$$

The barrier height depends on the width of the depletion layer (d), donor concentration ($N_D$) due to oxygen vacancies in the grains and the dielectric constant ($\varepsilon_r$). The resistance per square centimeter of the barrier is proportional to exp ($\phi$ / KT) [11,12,13]. If $\phi$ is small, PTCR jump would also be small. Further studies are in progress to modify the barrier height ($\phi$) in PFN-xPT ceramics so as to get larger PTCR jump.

**CONCLUSIONS:**

A high value of dielectric constant is obtained at lower frequencies in the PFN-0.13PT ceramics. The extremely high dielectric constant at low frequencies is shown to be due to the formation of barrier layers. The temperature dependence of resistivity shows PTCR behaviour due to the creation of oxygen vacancies in the grain during sintering at high temperatures.

**ACKNOWLEDGEMENT:**

It is a pleasure to acknowledge the interactions with Professor D. Kumar, Professor Om Parkash and Dr. H. Sharma.